\documentclass[iop]{emulateapj}
\usepackage{amsmath}
\usepackage{epstopdf}
\usepackage{multirow}
\usepackage{subfigure}
\usepackage{rotating}
\usepackage{longtable}

\newcommand{\mearth}{M$_{\oplus}$}
\newcommand{\rearth}{R$_{\oplus}$}
\newcommand{\msun}{M$_{\odot}$}

\newcommand{\teff}{$T_{\rm eff}$}
\newcommand{\kep}{{\it Kepler}}

\newcommand{\osix}{$^{16}${\rm O}}
\newcommand{\oseven}{$^{17}${\rm O}}
\newcommand{\oeight}{$^{18}${\rm O}}
\newcommand{\ctwo}{$^{12}${\rm C}}
\newcommand{\cthree}{$^{13}${\rm C}}

\slugcomment{ApJ, in press}

\shorttitle{Stellar Abundances and the Building Blocks of Planets}
\shortauthors{Gaidos}

\begin{document}


\title{What are Little Worlds Made Of?\\  Stellar Abundances and the Building Blocks of Planets}

\author{Eric Gaidos \altaffilmark{1}} 
\affil{Department of Geology \& Geophysics, University of Hawai`i at M\={a}noa, Honolulu, HI 96822}
\email{gaidos@hawaii.edu}

\altaffiltext{1}{Visiting Researcher, Max Planck Institut f\"{u}r Astronomie, Heidelberg, Germany}

\begin{abstract}

If the photospheres of solar-type stars represent the composition of
circumstellar disks from which any planets formed, spectroscopic
determinations of stellar elemental abundances offer information on
the composition of those planets, including smaller, rocky planets.
In particular, the C/O ratio is proposed to be a key determinant of
the composition of solids that condense from disk gas and are
incorporated into planets.  Also, planets may leave chemical
signatures on the photospheres of their host stars by sequestering
heavy elements, or by being accreted by the stars.  The presence,
absence, and composition of planets could be revealed by small
differences in the relative abundances between stars.  I critically
examine these scenarios and show that (i) a model of Galactic chemical
evolution predicts that the C/O ratio is expected to be close to the
solar value and vary little between dwarf stars in the solar
neighborhood; (ii) spectroscopic surveys of M dwarf stars limit the
occurrence of stars with C/O $\gtrsim 1$ to $<10^{-3}$; and (iii)
planetesimal chemistry will be controlled by the composition of
oxygen-rich dust inherited from the molecular cloud and processed in a
dust-rich environment, not a gas with the stellar composition.  A
second generation of more reduced planetesimals could be produced by
re-equilibration of material with dust-depleted gas.  Finally, I
discuss how minor differences in relative abundances between stars
that correlate with condensation temperature can be explained by
dust-gas segregation, perhaps in circumstellar disks, rather than
planet formation.
 
\end{abstract}
\keywords{stars:abundances -- planets and satellites:gaseous planets
  -- planets and satellites:formation -- methods:spectroscopic}

\section{Introduction}

Ground-based Doppler radial velocity surveys and space-based
photometric surveys have established that planets are very common, and
perhaps ubiquitous, around Sun-like stars.  In particular, analysis of
detections by the NASA \kep{} mission show that planets with radii of
1-2 Earth radii (\rearth{}) are far more common than larger planets
\citep{Petigura2013,Silburt2014}.  Comparisons between estimates or
upper limits on mass from Doppler radial velocity measurements and
radii from \kep{} also suggest that the densities of planets (at least
planets on short-period orbts) smaller than 1.5\mearth{} are
consistent with a ``rocky'' (silicates + metals) composition, while
larger planets have an additional low-molecular weight (e.g. hydrogen)
envelope \citep{Marcy2014}.  However, the data are insufficiently
precise to inform as to the exact composition of these planets and
regardless there are degeneracies in mass and radius with different
bulk compositions.

The photospheres of dwarf stars on the main sequence are broadly
representative of the material from which the star and any
circumstellar disk formed.  Additional information about planet
composition might therefore be inferred by determining elemental
abundances in the host-star photosphere.  This connection is supported
by the observation that the abundances of refractory elements in the
solar photosphere are, to a large extent, mirrored by that in
primitive chondritic meteorites, the latter widely used as an analog
of the primordial building blocks of planets
\citep[e.g.,][]{Lodders2003}.  To a lesser extent, these abundances
are reflected in the bulk composition of the planets such as Earth,
but with some important exceptions.

The two most abundant heavy elements in the Galaxy and Sun (and
presumably planet-forming disks) are carbon (C) and oxygen (O).  In
the cool interstellar medium (ISM), star-forming regions, and molecular
cloud cores, C and O are present primarily as the CO molecule
\citep[e.g.,][]{Bolatto2013}.  Because of the unit stochiometry of the
CO molecule, whether C or O is in excess controls which element is
available to form other compounds and thus the chemistry of the gas
and solids that condense or equilibrate with the gas.  The solar ratio
of these two elements is presently estimated at $0.55 \pm 0.12$
\citep[][see also Asplund et al. 2009]{Caffau2011}, however this value
is not necessarily universal.  The two elements differ in their
predicted pathways of stellar nucleosynthesis and C/O has a strong
positive correlation with metallicity \citep{Teske2014,Nissen2014}.

Motivated by condensation scenarios for the solar nebula
\citep{Barshay1976} and observations of widely varying C/O among
solar-type stars \citep{Edvardsson1993}, \citet{Gaidos2000} proposed
that C/O controls the abundance of water in planetary systems.  During
condensation from a hot gas with C/O $\ll 1$, excess oxygen forms
refractory silicate minerals and eventually H$_2$O.  However, in
systems with C/O near a critical value (about 0.88), no O for water is
available after silicate condensation and planets accrete from ``dry''
material.  At still higher C/O, the deficiency in O causes carbides
and graphite to become more stable and replace silicates in the
condensation sequence; thus ``carbide planets'' with very
un-Earth-like mineralogies, interior structures, and atmospheres might
form around stars with C/O $\gtrsim 1$.  

Many analyses of the spectra of solar-type stars obtained for galactic
chemical evolution (GCE) studies \citep{Gustafsson1999} and Doppler
radial velocity surveys found that a significant fraction have C/O
$\gtrsim 1$
\citep{Ecuvillon2004,Ecuvillon2006,DelgadoMena2010,Petigura2011}.  A
series of subsequent works have predicted the final bulk composition
of planets by modeling (i) the sequential condensation of elements and
formation of planetesimals from a cooling gas disk with an initial
composition set by host star abundances; and (ii) the accretion of
those planetesimals into planets via integration of the dynamical
equations of motion.  These scenarios have been applied to specific
systems where the photospheric abundances of elements have been
estimated, variously predicting water-rich ``ocean planets'' and
``carbon planets'' lacking any water
\citep{BondJ2010,Elser2012,Bond2012,Moriarty2014}.

One question at the foundation of these works is whether C/O really
varies widely among Galactic disk stars in the solar neighborhood and
ever approaches unity.  Abundances of C and O are more uncertain than
for many other elements because of the limited number of usable
absorption lines, confusion with lines of other elements, and non-LTE
corrections.  Recent studies have revised the C/O of nearby solar-type
stars, including those that host known planets, downwards
\citep{Nissen2013,Teske2013,Teske2013b,Teske2014}.  These newer
studies find no cases where C/O $>0.8$, the threshold where carbide
minerals are expected to first form.  \citet{Fortney2012} has pointed
out that the occurrence of C-rich systems in surveys of solar-type
stars conflicts with the lack of known C-rich ultracool T dwarfs, as
the latter are easily distinguished by their spectra
\citep{Fortney2012}.

A second issue is the assumption that the solids in planetary systems
condense from a cooling gas of stellar composition in a circumstellar
disk.  Equilibrium condensation can explain some solids in primitive
chondrites, widely considered analogs to the building blocks of
planets, but these solids constitute only a small fraction of those
meteorites.  Calcium aluminum inclusions (CAIs) and amoeboid olivene
agregates, the most abundant refractory condensates in chondritic
meteorites, have volume abundances $<10$\% and typically $<1$\%
\citep{Scott2007,Hezel2008}.  The dominant constituents of chondritic
meteorites are chondrules and fine-grained matrix, and there is
compelling evidence that these are the product of incomplete
sublimation and chemical alteration of pre-existing solids, e.g. older
generations of solids and even pre-solar dust from the parent
molecular cloud \citep{Huss2003,Trinquier2009,Burkhardt2012}.  This
processing occurred under conditions that were very different from a
gas of solar composition, i.e. an oxygen-rich environment and
dust-to-gas ratios $\ge 1$ \citep{Alexander2012}.  This environment
could have been created by gravitational settling to the mid-plane of
the protoplanetary disk.  Because of this, the composition of
planeteismals reflects the composition of dust grains in the molecular
clouds cores that collapse to form stars and planets, plus subsequent
thermal processing at mid-plane conditions.  ISM dust was, in turn,
the product of processing in the interstellar medium involving cycling
of order $10^2$ times between denser, cooler clouds, where it accreted
mantles of atoms, and the hotter intercloud phase, where shock-heated
ions and high-energy radiation (UV, X-rays, cosmic rays) sputtered
atoms from the grains \citep{Thielens2012}.

An equally intriguing possibility is that the compositions of
planetesimals and/or planets can be {\it inferred} from differential
measurements of abundances in the photospheres of stars.  This
approach is motivated by the detection of small differences in stellar
abundances of elements that are more refractory (higher condensation
temperature $T_c$) and thus more likely to be incorporated into
planets \citep[e.g.,][]{Melendez2009}.  These differences could arise
either from the accretion of planetesimals or planets onto the star,
after the dissipation of disk gas, or the sequestration of solids into
planets, and accretion of the dust-poor gas onto the
star.\footnote{This was one explanation for the well-established
  correlation between overall metallicity and giant planets
  \citep{Gonzalez1998,Santos2004,Fischer2005}}  This leads to a
prediction that these differences correlate with the presence or
properties of planetary systems, and provides a potential
``short-cut'' to discovering planets as well as estimating their
chemical composition.  However, subsequent analyses of larger samples
of stars does not support such a correlation
\citep{Gonzalez2013,Liu2014}, and alternative explanations for these
differences should be considered.

In this work, I critically examine the hypothesis that C/O varies
significantly among neighboring stars in the Galactic disk (and hence
the host stars of Doppler-detected exoplanets) from both theoretical
(Sec. \ref{sec.gce}), observational (Sec. \ref{sec.mdwarfs}), and
cosmochemical (Sec. \ref{sec.grains}) perspectives.  For the first I
develop a GCE model to predict the abundance of C and O over the
history of the Galactic disk in the vicinity of the Sun.  For the
second, I use the spectra obtained in M dwarf surveys to place strict
upper limits on the occurrence of C-rich single M dwarfs among recent
large surveys.  For the third, I combine a simple model of
interestellar dust evolution with UV measurements of the depletion of
heavy elements in the ISM to determine what controls the C/O ratio and
by how much it may vary.  Finally, I develop an alternative
explanation that can explain the observed elemental relative abundance
differences between solar-type stars, without resorting to possible
signatures of planet formation (Sec.  \ref{sec.gasdust}).

\section{Expectations for C/O from Galactic Chemical Evolution}
\label{sec.gce}

Galactic chemical evolution (GCE) models predict changes in the
abundances of elements and isotopes with time in both stars and the
ISM \citep{Prantzos2008}.  They combine yields of nucleosynthetic
products in stellar winds and ejecta with the stellar initial mass
function (IMF) and a prescription for star formation rate to estimate
the production of elements and isotopes.  These are integrated with
descriptions of the flow of heavy elements back into the ISM and
subsequent incorporation in long-lived low-mass stars and remnants
from high-mass stars.  Production varies with both stellar progenitor
mass and metallicity, thus the changing metallicity of new generations
of stars must be tracked in a GCE.

Carbon and oxygen are synthesized in stars more massive than
1.5\msun{} and injected into the interstellar medium through winds and
Type II supernovae (SN): Type I SN are predicted to contribute
$<0.3$\% \citep{Gehrz1998}.  GCE models generally predict a positive
trend of C/O with metallicity (and hence time) as well as variation
between stellar populations with different chemical histories
\citep[e.g.,][]{Tinsley1980,Timmes1995,Cescutti2009}.  These trends
are broadly observed within disk stars and between the thin,
thick-disk and bulge populations, and even between massive and
metal-poor irregular dwarf galaxies
\citep{Garnett1995,Cescutti2009,Nissen2014,Esteban2014}.  At least two
effects are thought to contribute to this correlation: (i) the short
main-sequence lifetime of massive stars, which are major contributors
to $\alpha$-elements including C and O, compared to intermediate mass
asymptotic giant branch (AGB) progenitors, which contribute
comparatively more to the C budget\footnote{While O yields are
  relatively insensitive to model parameters and are consistent from
  model to model \citep{Woosley1995}, the C yield of intermediate-mass
  stars depend sensitively on the amount of convective ``dredge-up''
  \citep{Renzini1981}.}; (ii) increased mass loss and C yield from
younger, more metal-rich AGB stars at the expense of O.  Mass-loss
rates depends on metallicity through its effect on the opacity of
outer stellar atmospheres and the impact on their structure.  The
implications of the abundance dependence of C/O yields for the
chemical evolution of galaxies has been previously pointed out
\citep[e.g.,][]{Maeder1992,Frayer1997}.

The C/O of bulge and disk populations diverge due to a combination of
differences in the stellar IMF, wind-driven mass loss in the former
and the accretion of primordial metal-poor gas in the former. The
flatter IMF of bulge stars implies a higher relative production of O
from more abundant massive stars. Addition (loss) of mass from a
star-forming system will increase (decrease) the amount of star
formation required to arrive at a given metal abundance and therefore
the C/O at that abundance. 

One feature of the statistics of stellar metallicities of the Galactic
disk in the neighborhood of the Sun is the relatively narrow
dispersion and paucity of metal-poor stars.  This is often called the
``G dwarf problem'' and translates into a flat age-metallicity
relation for the most of the history of the Galactic disk, at least in
the solar neighborhood.  It is actually not a problem for GCE models
if addition of metal-poor gas to the disk is admitted to balance
stellar nucleosynthesis \citep[e.g.,][]{Holmberg2007}.  If C/O is
strongly correlated with metallicity then a flat age-metallicity
relation might mean that C/O evolves very little
\citep[e.g.,][]{Wheeler1989}.

To re-visit the question of the evolution of C/O in the solar
neighborhood, the abundances of the two elements were calculated on an
isotope-by-isotope basis using the model described in detail in the
Appendix.  The model simulates the production of these isotopes and
release into the ISM in SN explosions, winds from massive stars and
AGB stars, sequestration of isotopes into long-lived low-mass stars,
and addition of metal-poor gas by infall onto the Galactic disk.  The
ISM is described by two components; an inter-cloud medium which
produces molecular clouds but no stars, and a star-forming giant
molecular cloud component.  Yields for intermediate- and massive stars
are taken from a variety of recent sources.  Best-fit parameters
describing the timescale of exponentially-declining gas infall on the
disk, the initial metallicity of the infalling gas, the age of the
stellar population at the solar galactocentric radius, the index of
the power-law describing the IMF for massive stars, and the power-law
relationship between star formation rate and gas surface density were
found by a Monte-Carlo Markov Chain analysis: The observational
constraints were the age of the Sun, the present mass surface density
of stars, stellar remanants, and gas at the solar galactocentric
radius, solar metallicity (here taken to be C+O), the current
metallicity of the ISM, and its intrinsic standard deviation.

\begin{figure}
\includegraphics[width=84mm]{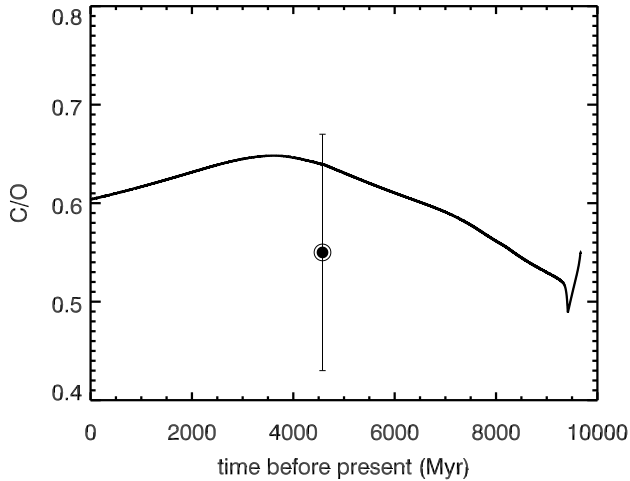}
\caption{Evolution of C/O at the solar galactocentric radius according
  to the GCE model described in the text and Appendix.  The solar
  value and its uncertainty from \citet{Caffau2011} is plotted.}
 \label{fig.co}
\end{figure}

The prediction evolution of C/O is compared to the solar value from
\citet{Caffau2011} in Fig. \ref{fig.co}.  After initial transients
that die away after a few hundred Myr, there is only moderate
evolution in C/O during most of the history of the Galactic disk.  The
C/O ratio rises from 0.5 to about 0.65 over the first 6~Gyr in
response to increasing input from stars of lower mass and higher
metallicity, as described above, and declines slightly thereafter.
The predicted value at the time of the Sun's formation is 0.64,
withinin $1\sigma$ of the \citet{Caffau2011} value.  The formal error
in the predicted value, based on the standard deviation of the MCMC
chain after removal of the ``burn in'', is only 0.003, but this value
does not reflect the dominant source of uncertainty in these
calculations -- the nucleosynthetic yields.  Although the model is
unable to exactly reproduce the solar values of C and O abundances,
correctly reproducing these using GCE been always been challenging.
\citet{Maeder1992} also found that satisfactorily reproducing a
``flat'' C/O places constraints on the remnant or ``cut-off'' mass at
the center of a SN progenitor which is not injected into the ISM.

\begin{figure}
\includegraphics[width=84mm]{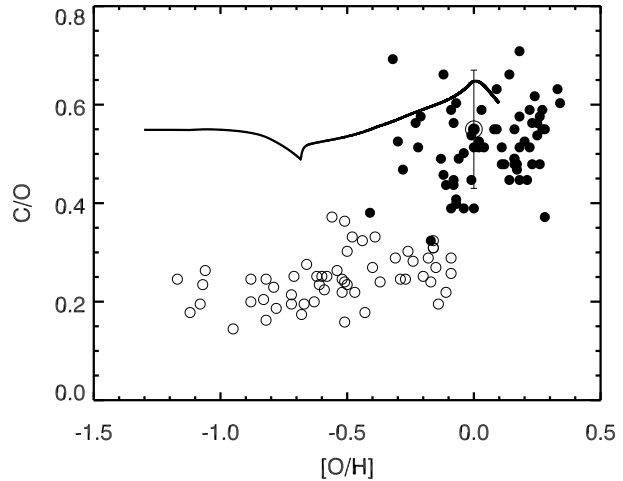}
\caption{Evolution of C/O at the solar galactocentric radius according
  to the GCE model described in the text and Appendix, plotted
  vs. [O/H].  The \citet{Caffau2011} solar value is plotted as well as
  values for Galactic disk stars (filled points) and bulge/halo stars
  (open points) from \citet{Nissen2014} and tied to the solar value.}
 \label{fig.co2}
\end{figure}

Figure \ref{fig.co2} plots the predicted C/O evolution vs. predicted
[O/H], again compared to the Sun.  Also plotted are the values found
by \citet{Nissen2014} for Galactic disk stars (filled points) and
bulge or halo stars (open points) and converted to absolute values
assuming a solar C/O of 0.55.  The predicted trend is perfectly
consistent with the disk values if an offset of about 16\% with the
predicted vs. measured Solar value is artificially removed.  The
absence of data for disk stars at low [O/H] is a consequence of the
age-metallicity relationship: comparatively little time elapsed
($<1$~Gyr according to the model) and few stars formed in this
interval.

The predicted constancy of C/O contrasts with previous predictions for
an increase in C/O with metallicity/time.  For example, both
\citet{Cescutti2009} and \citet{Mattsson2010} predict a positive slope
of $\sim 1$ dex/dex in [C/O] vs. [O/H], and thus a C/O of about 0.06
at [O/H] = -1.  Since the model and the best-fit parameter values
presented here are similar to those of previous works the most likely
cause of this difference is in the particular nucleosynthetic yields
that were used.  The yields from AGB stars have been significantly
revised \citep{Karakas2010} but in the model these are minority
contributors to the C and O budgets.  Indeed, if there contribution is
completely removed, C/O still changes little and the predicted solar
C/O is slightly closer to the \citet{Caffau2011} value.  Instead, the
predictions of this model probably stand out from previous results
because of the dominant contribution of massive stars to both C and O.
Since these stars have very short lives compared to the chemical
evolution of the disk, there is no effect from delayed contribution,
and the C/O rapidly approaches a steady-state value.

Previous works have pointed to favorable comparisons of model
predictions with observations, but the latter are {\it combined} data
on bulge, halo, and thick- and thin-disk populations, therefore
assuming a common origin for these populations that may not exist.
For example, the [C/O] of thin disk stars presented by
\citet{Bensby2008} and \citet{Cescutti2009} show no correlation with
      [O/H] and it is only when bulge and thick disk stars are added
      that such a correlation appears.  The apparent consistency
      between the predicted and observed trends of [C/O] vs. [O/H]
      \citep{Cescutti2009,Mattsson2010} belies the fact that the
      models were tuned for the (thin) disk population in the solar
      neighborhood and should not be compared with other stellar
      populations.

It is indisputable that metal-poor bulge/halo stars have low C/O
compared to the solar neighborhood \citep{Fabbian2009,Nissen2014}.
The model presented here cannot, and was not designed to, explain
these populations, but their abundances might be a result of a flatter
IMF, early loss of gas and/or truncation of star formation.  But
bulge/halo stars are rare in the solar neighborhood and because their
lines are very weak they are usually avoided by planet searches using
the Doppler radial velocity method.

C/O may also vary at some level within the solar galactocentric
annulus because the sources of nuclides (stars and stellar clusters)
are discrete and mixing by rotational shear and epicyclic dynamics is
not completely efficient.  However, GCE is occurring on a timescale
much longer (few Gyr) than the rotational time of the Galaxy at 8~kpc
($\sim$250~Myr) which governs the rate of mixing, and thus this limits
the magnitude of such heterogeneities.

\section{M Dwarfs as Probes of High C/O}
\label{sec.mdwarfs}

In solar-type stars, the absorption lines of C and O are weak and C/O
has little effect on the overall spectrum.  \citet{Fortney2012}
pointed out that cooler stars, i.e. brown dwarfs, offer more obvious
constraints on the occurrence of high C/O systems.  The spectra of M
dwarfs, with effective temperatures \teff{} $<3900$K, also include
absorption bands of O-containing molecules, e.g. TiO, VO, and CaOH,
which are sensitive to the available O abundance \citep{Schmidt2009},
and hence C/O.  At C/O $\ge 1$, TiO is absent and the Schwan bands of
C$_2$, CN, and CH should appear.  Such spectra are characteristic of
carbon stars, evolved stars in which dredge-up of carbon-righ
interiors has occurred (AGB stars).  If the initial mass function of
star formation is chemically invariant, a limit on the occurrence of
C-rich M dwarfs is also a limit on C-rich G dwarfs.

Dwarf carbon (dC) star have been identified \citep{Dahn1977} but these
probably accreted carbon-rich gas from a present or former evolved
companion \citep{Behara2010,Green2013}.  Overall, dC stars are
uncommon \citep{Downes2004,Green2013}.  Carbon enhanced metal-poor
(CEMP) stars in the Galactic halo appear are also rare overall but
more prevalent among (initially) metal-poor systems because less
accreted C is required to increase C/O \citep{Beers1992}.  Any tally
of {\it intrinsically} carbon-rich dwarfs must remove these
interlopers or be considered an upper limit.

I considered three spectral indices which are continuum-normalized
measures of the emission in specific bands \citep{Reid1995}: CaH,
which is the mean of the CaH2 (6814-6846\AA) and CaH3 (6960-6990\AA)
indices; TiO5 (7126-7135\AA); and CaOH (6230-6240\AA).  Values of
these temperature- and gravity-sensitive indices are highly and
positively correlated among solar-metallicity M dwarfs.  Carbon-rich
stars can be identified by relatively weak bands of oxygen-containing
TiO and CaOH, i.e. high TiO5 or CaOH indices, for a given strength of
the non-oxygen-containing CaH.  However, indices of metal-poor
subdwarfs or ``extreme'' subdwarfs exhibit similar behavior
\citep{Jao2008,Woolf2009,Rajpurohit2014}.

\begin{figure}
\includegraphics[width=84mm]{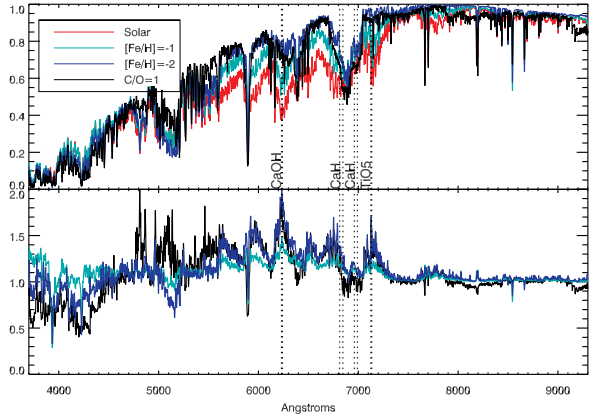}
\caption{Top: Normalized PHOENIX BT-Settl synthetic spectra of M
  dwarfs with $T_{\rm eff} = 3700$~K, $\log g = 5$, and solar
  abundances \citep[][ red line]{Caffau2011}, metal-poor subdwarf
  ([Fe/H] = -1, [$\alpha$/Fe] = 0, green line), extreme subdwarf
  ([Fe/H] = -2, [$\alpha$/Fe] = +0.4, $\log g = 5.5$, blue line), and
  solar-metallicity star except some O as C such that C/O=1 (red
  line).  Bottom: ratios of the three non solar-metallicity spectra
  described above to the solar metallicity spectrum.}
 \label{fig.spectra}
\end{figure}

To guide discrimination between C-rich M dwarfs and metal-poor stars
based on indices, synthetic spectra were generated using the BT-Settl
version of the PHOENIX models \citep{Allard2011}.  Four cases were
considered: a solar-metallicity star, a star with [M/H] = 0 but C and
O adjusted so that C/O=1, a metal-poor subdwarf (sd) with [M/H] = -1,
and an extreme subdwarf (esd) with [Fe/H] = -2.  The last two cases
also had an alpha-element enhancement [$\alpha$/Fe] = +0.4.  Effective
temperatures over the entire M dwarf and late-K dwarf range in steps
of 100~K were considered.  All cases had $\log g = 5$, except for the
esd which had $\log g = 5.5$.  The visible and far-red portions of the
spectra are plotted and compared in Fig. \ref{fig.spectra}.  The bands
used to compute the CaH, TiO5, and CaOH indices are indicated.  The
current BT-SETTL line lists do not actually include CaOH, but the
bandpass also includes the TiO $\gamma$ line which is also sensitive
to C/O.

As expected, in the C/O=1 case (black lines in Fig. \ref{fig.spectra})
the TiO bands at around 6235 and 6700\AA{} are dramatically weakened
relative to the solar case (red line) while those of CaH are
essentially unchanged.  This is also true for the esd case (but not
the subdwarf case).  The differential response of TiO5 and CaH is the
basis for the ``$\zeta$'' parameter developed by \citet{Lepine2007}.
However, the C/O=1 and esd cases differ markedly at 4700-5300\AA{} and
below 4500\AA{} because Fe lines are very weak in the esd case.  There
are also difference at $\lambda > 8000$\AA{} (Fig. \ref{fig.spectra}).
Absorption by H$_2$O over a broad wavelength range centered at
1.9$\mu$m is also weaker in the C/O=1 case, but this would be
difficult to ascertain from the ground.  Index values were calculated
for the four cases at each \teff{}.

\begin{figure}
\includegraphics[width=84mm]{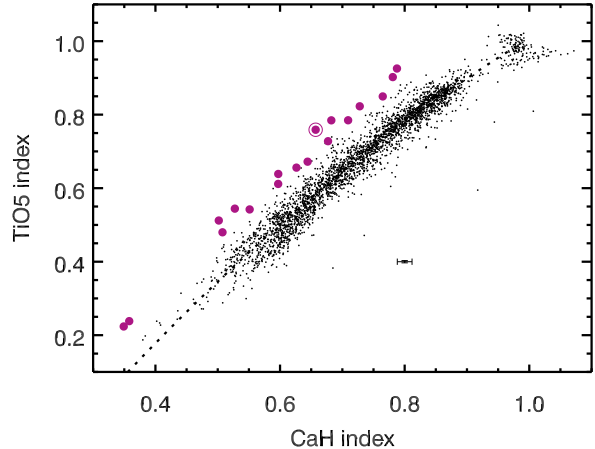}
\caption{CaH vs. TiO5 indices for 2583 stars from the CONCH-SHELL
  catalog of \citet{Gaidos2014}, some with repeated observations.  The
  black dashed curve is a quadratic fit to the locus.  Eighteen stars
  with anomalously weak TiO5 bands (large indices) for their CaH band
  strength are marked as magenta points.  The circled point (Wolf~1130
  or GJ~781) has the most deviatory TiO5 index.  The isolated point
  illustrates the median errors.}
 \label{fig.indices}
\end{figure}

I searched for carbon-rich M dwarfs in three spectroscopic catalogs:
the CONCH-SHELL spectroscopic catalog of nearby, bright M dwarfs
\citep{Gaidos2014}, the Sloan Digital Sky Survey (SDSS) catalog of M
dwarfs constructed by \citet{West2011}, and the LAMOST spectroscopic
catalog of M dwarfs in \citet{Yi2014}.  Stars in the CONCH-SHELL
catalog were selected based on brightness ($J < 9$), parallaxes or
proper motions consistent with main sequence status, and colors.
Although some sets of stars in the catalog were selected based on
optical and infrared colors consistent with ``normal'' M dwarfs, this
criterion was relaxed for another set, thus admitting any C-rich
dwarfs with peculiar colors.  The selection for ``red'' ($V-J > 2.7$)
stars is not relaxed, and this inevitably eliminates some metal-poor
early M-type dwarfs, although not necessarily C-rich examples.  Figure
\ref{fig.indices} plots the TiO5 vs. CaH indices for CONCHSHELL stars.
The red line is the best-fit quadratic to the locus and the black
points and lines are the values computed from PHOENIX BT-SETTL
spectra.

\begin{figure}
\includegraphics[width=84mm]{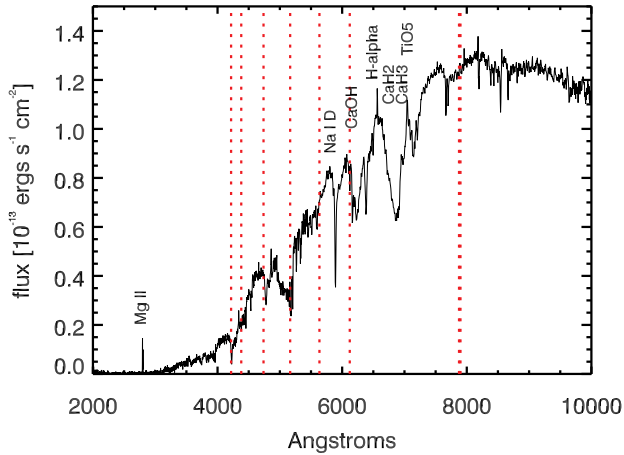}
\caption{HST STIS spectrum of Wolf~1130/GJ~781, the star with the
  comparatively weakest TiO5 band in the catalog (circled point in
  Fig. \ref{fig.indices}).  Some atomic lines and molecular bands are
  labeled; the red dashed lines mark the locations of some CN and
  C$_2$ Schwan bands observed in carbon stars, but not in this star.}
 \label{fig.stis}
\end{figure}

I identified 18 CONCH-SHELL stars with a TiO5 band that is
significantly ($>3\sigma$) weaker (larger index) compared to a
best-fit second-order fit of TiO5 as a function of CaH.  When
calculating the significance of a deviation, an intrinsic dispersion
of 0.032 in TiO5 was added in quadrature to the formal measurement
errors.  None of these 18 stars had indices that deviate to the extent
predicted for a C/O=1 star.  The most deviant star is
PM~I20050+5426/GJ~781/Wolf~1130, with a TiO5 index that is 0.17 above
the best-fit locus value.  This was previously identified as an active
M1.5 subdwarf \citep{Gizis1997} with a metallicity based on an
infrared spectrum of $-0.64 \pm 0.17$ \citep{Rojas-Ayala2012}.
Interestingly, \citet{Gizis1998} identified Wolf~1130 as a
single-lined spectroscopic binary ($P \approx 0.5$~d) and proposed
that the unseen companion was a 0.3\msun{} helium white dwarf.  A
Hubble Space Telescope Imaging Spectrograph (STIS) spectrum of the
star (Fig. \ref{fig.stis}) exhibits some TiO absorption and no
diagnostic carbon-star features.  There is also no indication of a WD
companion, although this would be consistent with an advanced age and
hence low UV luminosity of any such object.  Spectra of the other
candidates indicate they are metal-poor stars, or have systematic
errors.

\begin{figure}
\includegraphics[width=84mm]{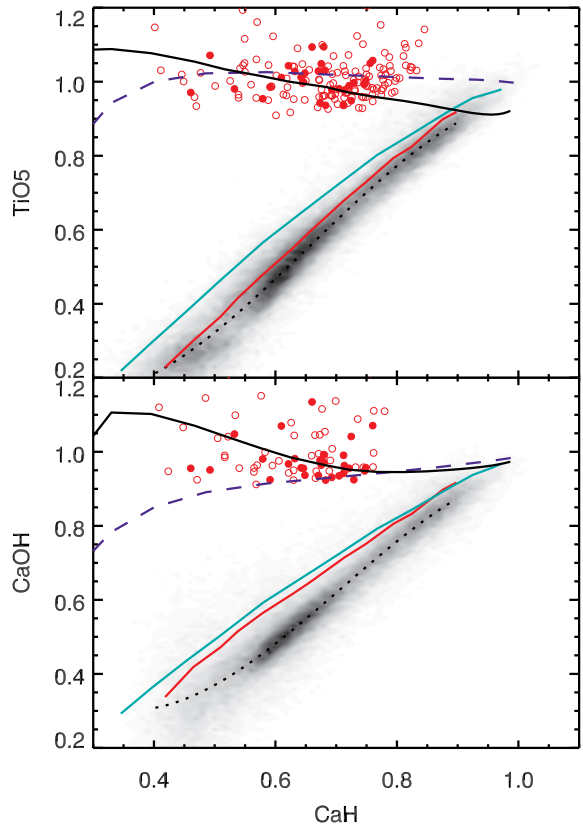}
\caption{CaH, TiO5, and CaOH indices of M dwarfs in Data Release 7 of
  the SDSS \citep{West2011}.  The density of stars is shown as a grey
  scale.  The dashed black curves are the best fits to the loci.  The
  red, blue, purple dashed, and black solid curves are the predictions
  of the PHOENIX BT-SETTL model for solar-metallicity, subdwarf,
  extreme subdwarf, and C/O=1 cases, respectively.}
 \label{fig.dr7}
\end{figure}

I examined the molecular indices for two much larger samples of M
dwarfs from the SDSS and LAMOST \citep{West2011,Yi2014}.  Figure
\ref{fig.dr7} shows the TiO5 and CaOH indices vs. CaH index for 70,841
M dwarfs with spectra in Data Release 7 of the Sloan Digital Sky
Survey \citep{West2011}.  The grey scale is linearly related to the
density of stars in the TiO5- or CaOH-vs. CaH plots.  The black dashed
line is the best-fit to the stellar locus, and the red, aquamarine,
dashed purple, and black solid lines are the predictions from the
BT-SETTL models for the solar metallicity, subdwarf, extreme subdwarf,
and C/O=1 cases, respectively.  Stars for which either index value is
significantly ($>3\sigma$) higher than the best-fit polynomial to the
stellar locus, and within $1\sigma$ of or above the predicted C/O=1
locus and are plotted as open points.  The 30 stars where both indices
satisfy these criteria are plotted as filled points.  Inspection of
the SDSS spectra of these 30 stars found that all are consistent with
template spectra of solar-metallicity or metal-poor stars and do not
have the featuers expected of C/O=1 stellar atmospheres.

\begin{figure}
\includegraphics[width=84mm]{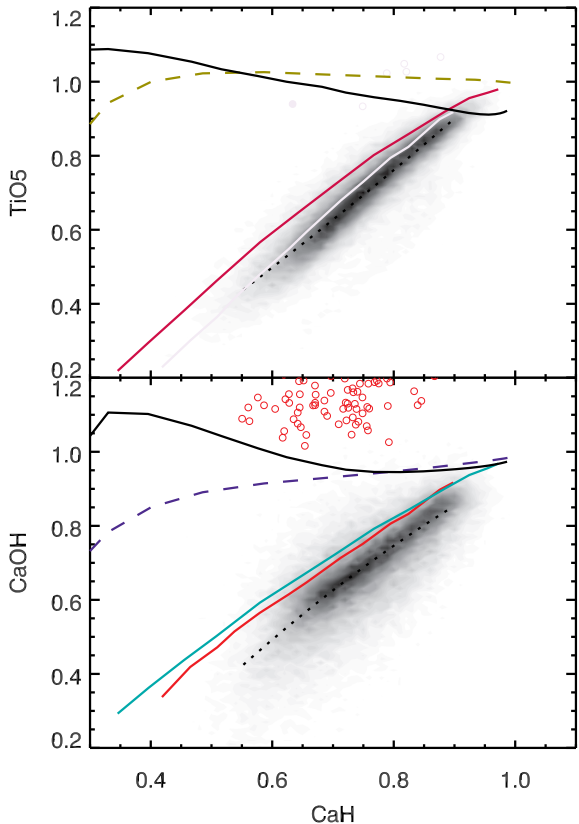}
\caption{Same as Figure 6, but for candidate M dwarfs from the LAMOST
  sample.  Several stars with anomalous index values lie above the
  plots.}
 \label{fig.lamost}
\end{figure}

Figure \ref{fig.lamost} is the analogous set of plots for 67,082
candidate M dwarfs from a pilot survey of the Large Area Multi-Object
Fiber Spectroscopic Telescope \citep[LAMOST,][]{Yi2014}.  The
intrinsic scatter in index values, after correcting for measurement
errors, is much larger for this sample, perhaps due to systematic
errors.  Only two stars have both TiO5 and CaOH values significantly
above the locus, and within 1$\sigma$ of the predicted C/O line.
However, the index values of these stars are all $\gg 1$ and probably
spurious.  The paucity of high C/O candidates in the LAMOST survey
compared to the SDSS DR7 sample is undoubtedly due to the larger
uncertainties in the indices.

Based on the CONCH-SHELL sample alone, high C/O $\sim 1$ stars
constitute less than $1.2 \times 10^{-3}$ of M dwarfs with 95\%
confidence, a stricter constraint than the limits placed by
\citet{Nissen2014} based on spectroscopy of 66 solar-type stars.  The
null result from the DR7 sample places an an even more stringent upper
limit of $6 \times 10^{-4}$ at 99\% confidence.  It is possible that
manual screening performed by \citet{West2011} to remove corrupted
spectra could have excluded carbon-rich stars, but at the resolution
and signal-to-noise of the spectra, the differences that distinguish
C/O=1 spectra from the solar case would not make them overwhelmingly
abberant.

\section{From Dust to Planetesimals}
\label{sec.grains}

Theoretical studies linking the abundances in the photospheres of host
stars to the composition of rocky planets generally assume that the
building blocks of planetesimals condensed directly from a cooling
gaseous disk of the same composition, but there is accumulating
meteoritic evidence that primitive solids in the Solar System were by
and large {\it not} produced by de nova condensation from a gas of
solar composition.  Instead, this material was the outcome of partial
thermal processing of pre-existing solids under high dust/gas ratios
and oxygen abundance (fugacity, $f_{O_2}$) brought about by growth and
settling of grains to the nebular mid-plane and/or inward migration of
water-rich (and hence O-rich) planetesimals \citep{Grossman2008}.
This evidence includes (i) the survival of pre-solar grains and
isotopic anomalies inherited from the molecular cloud and older
generations of stars \citep{Davis2011}; (ii) the gradual variation in
the abundance of elements with condensation temperature in primitive
meteorites, which contradicts the sharp cutoff predicted by
equilibrium condensation and which cannot be explained by radial
transport in the disk \citep{Ciesla2008}; (iii) the retention of
volatile sodium and sulfur and lack of isotopic mass fractionation for
potassium and silicon in chondrules \citep{Scott2007}; and (iv)
molybdenum and tungsten depletions in refractory inclusions
\citep{Fegley1985}.  The equilibrium oxygen fugacities of minerals in
many meteorites also suggest high ambient $f_{O_2}$, but some could
also be explained by alteration on the parent bodies of the meteorites
\citep{Grossman2008}.

Dust growth and settling to the midplane of a disk is predicted to
occur in $10^3$-$10^5$~yr, depending on the intensity of turbulence in
the disk \citep{Nomura2006,Ciesla2007}.  Dust settling, as well as
growth, might be observed via its effect on the spectral energy
distribution of a disk \citep{Tanaka2005}.  Tentative evidence for
significant dust settling and depletion from the upper layers of the
disks of T Tauri stars (ages of $\sim 10^6$~yr) has been presented
\citep{Furlan2006}, but unambiguous detection of settling is
challenging \citep{Murakawa2014}.  If settling occurs faster than the
viscous accretion time of a disk \citep[$10^6$~yr,][]{Hartmann1998}
then grains will experience high temperatures only in a dust-rich
environment.

If the precursor material of planetesimals is dust, rather than gas,
then to a large extent the composition of interstellar grains governs
the composition of rocky exoplanets.  Interstellar dust begins its
existence as condensates in the cooling envelopes and winds of AGB and
red giant branch stars but these grains are subsequently and
completely altered by many cycles of erosion and formation of mantles
in the ISM.  Erosion takes place in the lower-density,
higher-temperature inter-cloud phase of the ISM, principally by
sputtering by ions heated by the passage of supernova shocks as well
as UV photons.  Condensation takes place onto surviving grains that
are incorporated into the denser, cooler cloud phase of the ISM.  The
cycling time between these two phases ($\sim 3 \times 10^7$~yr, set by
the cloud lifetime) is short compared to the mean time since formation
in a circumstellar wind ($\sim 3 \times 10^9$~yr), hence the bulk
elemental composition of ISM grains is set by the balance between
condensation and sputtering \citep{Draine2003}.

The bulk composition of dust can be inferred by measuring the
depletion of elements from the gas phase, i.e. using the strength of
UV absorption lines along different lines of sight through the ISM to
some suitable background source.  Following \citet{Tielens1998}, the
equations of motion of the depletions $\delta$ in the cloud ($c$) and
inter-cloud ($i$) medium of any particular element can be written as:
\begin{equation}
\frac{\delta_c}{dt} = -k_2 \left( \delta_c - \delta_i \right) + k_4 \left(1 - \delta_c \right),
\end{equation}
and 
\begin{equation}
\frac{\delta_i}{dt} = -k_1 \left( \delta_i - \delta_c \right) - k_3 \delta_i,
\end{equation}
respectively.  In these equations, $k_1$, $k_2$, $k_3$, and $k_4$ are
the rate of mixing from the intercloud to the cloud medium, the rate
of mixing from the cloud to the intercloud medium, the rate of grain
destruction in the intercloud medium, and the rate of grain growth
from molecular cloud gas, respectively.  The nucleosynthetic rate of
production of an element is much slower than any of the rates of the
formation of clouds, dissipation of clouds, destruction by sputtering
in the inter-cloud phase, and growth condensation in the cloud phase,
respectively, and is ignored here.  The steady state solutions are:
\begin{equation}
\delta_c = \frac{1 + k_1/k_3}{1 + k_1/k_3 + k_2/k_4},
\end{equation}
and
\begin{equation}
\delta_i = \frac{k_1/k_3}{1 + k_1/k_3 + k_2/k_4}.
\end{equation}
Thus, the steady-state abundances are governed by only two parameters,
the amount of growth in the clouds $k_4/k_2$, and the amount of
erosion between clouds: $k_3/k_1$.

\begin{figure}
\includegraphics[width=84mm]{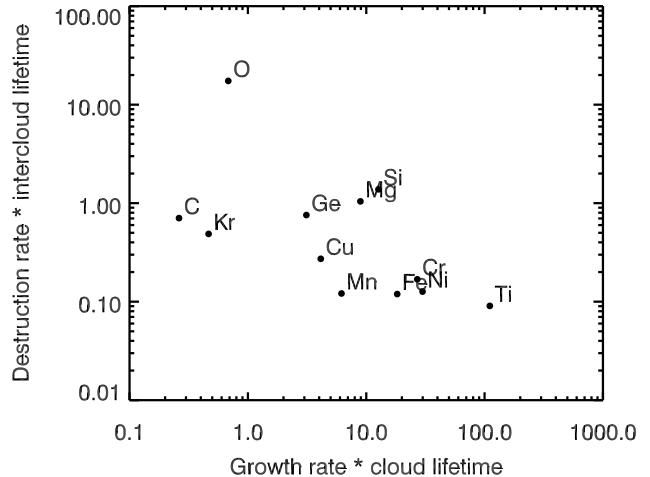}
\caption{Inferred rates of growth (in molecular clouds) and depletion
  (in the inter-cloud medium) of elements in interstellar grains based
  on the observations analyzed in \citet{JenkinsE2009}.  The rates are
  normalized by the residence time of grains in the cloud and
  inter-cloud media, respectively.}
 \label{fig.rates}
\end{figure}

I estimated parameter values for some elements using the data compiled
by \citet{JenkinsE2009}.  I set $\delta_c$ and $\delta_i$ to $1 -
[X/H]_1$ (maximum depletion) and $1 - [X/H]_0$ (minimum depletion),
respectively, and solved for $k_3/k_1$ and $k_4/k_2$.  Figure
\ref{fig.rates} plots these two parameters.  Three elements with data
in \citet{JenkinsE2009} are not shown: to explain the abundances of P,
Cl, and Zn in the context of this model requires {\it negative}
destruction in the intercloud medium.  This may be an artifact of
photoionization since the abundance of these elements is estimated
from their singly ionized forms.

The comparatively low condensation rates of Kr, C, and O reflect the
volatility of these elements.  The behavior of O deviates strongly
from that of the refractory elements, perhaps because the primary
carrier of O is not silicates but a much more volatile substance such
as water ice.  Likewise C does not behave as a refractory such as
graphite \citep{Tielens2012}.  This suggests that C is present as
relatively volatile organic matter \citep{Jones2009}.  Kr is not
expected to condense and the non-zero value of $k_4/k_2$ may be a
consequence of measurement errors or departures from solar relative
abundances \citep[]{Cartledge2003}.

The differences in the parameters for C and O manifest themselves as a
modified C/O ratio in interstellar dust with respect to the total
(gas+dust) abundance.  In the case of the nominal rates inferred from
the data of \citet{JenkinsE2009}, the degree of depletion from cold
molecular cloud gas shows that the C/O of the dust is 0.92 times that
of the bulk ISM, i.e. slightly more oxygen-rich than the current bulk
ISM value.  Although O is depleted more rapidly from interstellar
grains, it is also accreted more rapidly.  One caveat of this estimate
for the dust C/O is that the gas-phase depletion of C is more
uncertain than other elements because there are few suitable
absorption lines \citep{Sofia2009}.

The C/O could also vary between locations as a result of varying
$k_3/k_1$ and/or $k_4/k_2$.  \citet{Tielens1998} estimated the
residence time $1/k_2$ in the warm inter-cloud phase of the ISM as $3
\times 10^6$~yr assuming that it is set by the timescale for shocking
by SN and subsequent cooling and collapse into clouds.  This was
consistent with an ISM mass fraction in the warm phase of $\sim$10\%.
However, the mass in the warm phase is probably comparable to the
dense molecular H$_2$ (cloud) phase \citet{Draine2011} and thus the
residence time in the inter-cloud phase is similar to the cloud
lifetime $3 \times 10^7$~yr \citep{Murray2011}.  Over this time, dust
grains may experience $\sim10$ SN shocks before becoming incorporated
into molecular clouds.  In contrast, $k_4/k_2$ would be expected to
vary only to the extent that the lifetimes of molecular clouds varies.

\begin{figure}
\includegraphics[width=84mm]{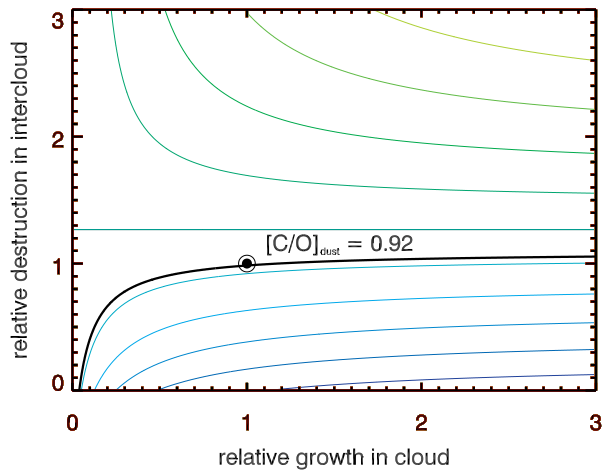}
\caption{Predicted variation of C/O in interstellar dust with rates of
  dust accretion and destruction relative to canonical parameter
  values, using the model of \citet{Tielens1998} and canonical values
  set by the observed depletions in the ISM.  Contours are intervals
  of 0.1, increasing upwards, and the canonical rates yield a C/O of
  0.92.  Rates for each element are assumed to vary by a uniform
  multiplicative factor.}
 \label{fig.contour}
\end{figure}

I estimated the sensitivity to variations in these rates by
multiplying each of the parameter ratios by varying factors $\in
[0,3]$.  Figure \ref{fig.contour} plots the predicted variation of C/O
with contours intervals of 10\%.  The unadjusted parameter values,
which predict [C/O]$_{\rm dust}=0.92$ (heavy contour), are at unit
value abcissa and ordinate (circle).  Variation in the efficiency of
grain destruction in the intercloud medium ($k_3/k_1$) have more
effect on dust C/O than variation in the efficiency of grain growth in
molecular clouds ($k_4/k_2$).  This difference is a consequence of the
larger dispersion (a ratio of $\sim 30$) in the removal of O vs. C
during grain destruction compared to the dispersion in the
incorporation of O vs. C during grain growth (a ratio of $\sim 3$,
Fig. \ref{fig.rates}).  The very volatile behavior of O compared to
the other elements is presumably because some of it is incorporated as
water ice mantles around dust grains.

This model predicts that large (factors of two) variation in dust
lifetime produced by different shock and UV conditions in the
intercloud medium will produce modest ($\sim$20\%) variation in bulk
dust C/O, with dust in the vicinity of massive SN progenitors more
carbon-rich.  By the arguments presented above, variation in dust C/O
could generate diversity in the composition of planeteismals, but this
will be limited by the extent that water ice is retained on dust
grains (see below).

\section{Elemental Abundance Variation among (Planet-Hosting) Stars}
\label{sec.gasdust}

A pronounced correlation between the metallicity of the host star and
the occurrence of giant planets was discovered early in the history of
exoplanet science \citep{Gonzalez1998}.  Increasingly precise
measurements of the relative abundances of elements in stellar
photospheres have permitted more nuanced investigations of
correlations with planet occurrence \citep[e.g.,][]{Robinson2006a}.
\citet{Melendez2009} found that the solar composition was anomalous
compared to 9 out of a sample of 11 solar ``twins'' and that the Sun
is deficient ($\le 0.1$~dex) in refractory elements (condensation
temperature $T_C > 900$~K) compared to more volatile elements.  They
proposed that this was a signature of the sequestration of those
elements in rocky planets and cores of gas giants.
\citet{Ramirez2010} also found a statistical correlation between
planets and a deficit of refractory elements among larger samples of
solar-type stars in Doppler surveys.  If differences in relative
photosphere abundances are a signature of rocky planets or giant
planet cores, this would provide a short-cut to planet detection and
even a means to probe the composition of the planets themselves
\citep{DelgadoMena2010}.

On the other hand, \citet{Gonzalez2013} find no clear correlation
between the slope of the relative abundance vs. $T_C$ and the presence
or absence of planets.  One factor that may contribute to these
contrasting results is the removal of galactic chemical evolution
trends, e.g. decreasing [$\alpha$/Fe] with increasing metallicity, by
\citet{Gonzalez2013}.  Moroever, \citet{Onehag2014} found no trend
with $T_C$ in the relative abundances of members of the solar-age,
solar-metallicity cluster M67.  Observations of physical pairs of
stars with and without (detected) planets provide the clearest test of
the planet-abundance relation, as both stars should have formed from
the same molecular cloud.  But the available observations do
unambiguously support such a correlation
\citep{TucciMaia2014,Liu2014}.

In fact, the trend of increasing differential abundance with
increasing condensation temperature approximately mirrors the pattern
of depletion of elements in ISM gas \citep{Yin2005}.  This simply
reflects the preferential condensation of refractory elements onto
interstellar grains and the universality of chemistry.  If
interstellar dust does not perfectly follow the gas, this will produce
variation in element abundances that are proportional to the extent of
gas-dust segregation and the dust composition.  All else being equal,
the abundances of the more refractory elements (which are mostly in
the dust) would be expected to vary more than volatile elements (which
are mostly in the gas), in a manner that roughly correlates with
$T_C$.

There is evidence for, and theoretical predictions of, dust-gas
segregation in the ISM.  \citet{Padoan2006} found that the power-law
describing the spatial distribution of extinction (a tracer of the
column density of dust) on the sky is shallower than that of $^{13}$CO
(a tracer of the gas column density), such that in the densest regions
of the ISM, extinction by dust is less than that predicted by perfect
correspondence with the gas.  This could be explained by grain growth
in excess of that predicted by models. \citep[CO condensation would
  presumably act in the opposite sense][]{Whittet2010}.  But
\citet{Padoan2006} offer spatial variation in dust-to-gas ratio as an
alternative explanation.

Dust-gas segregation could occur at three different scales; (i) over
the extent of a giant molecular cloud or star-forming region; (ii) in
cloud cores that collapse to form individual stellar systems; and
(iii) in the accretion disks around young stellar objects.
\citet{Draine2011} estimated that dust drift due to radiation pressure
from O stars can remove dust from the centers of H II regions in $\le
1$~Myr, but since such low-density regions are not themselves the site
of star formation, this effect should not manifest itself in the
relative abundances of stellar photospheres.

\citet{Bellan2008} showed that the different dynamics of dust and gas
during Bondi-type accretion flow can enhance dust-to-gas ratios by an
order of magnitude in cloud cores.  The mechanism modeled by
\citet{Bellan2008} relies on dust velocities of several km~sec$^{-1}$
relative to the gas around the growing cloud core.  The size
distribution of interstellar dust grains peaks sharply near 0.2-0.3
$\mu$m \citep{Draine2003}.  These grains may be dynamically decoupled
from the diffuse interstellar medium (number density $n \sim 0.1$
cm$^{-3}$, stopping distance $\sim 40$~pc), and there is tentative
evidence for this in the trajectories of interstellar meteors
\citep{Taylor1996} and {\it Ulysses} spacecraft measurements
\citep{Kruger2009}.  However, these grains will be tightly coupled to
the molecular gas ($n \sim 10^2-10^3$~cm$^{-3}$) surrounding a cloud
core, i.e. over length scales $\ll 1$~pc and much smaller than a
typical cloud size.  Under such conditions it seems unlikely that dust
acceleration mechanisms \citep[e.g.,]{Yan2009,Hoang2012} can achieve
equipartion of energy between ISM and dust and speeds of
$\sim$10~km~s$^{-1}$.

The second scenario which could produce variation in the dust-to-gas
ratio in cloud cores is drift induced by radiation pressure.
\citet{Whitworth2002} showed that the inward radial drift of
0.1-$\mu$m grains under the influence of a typical radiation field can
treble the dust concentration in a static gas sphere of a few solar
masses at 10~K in 10~Myr.  The effect scales linearly with the
intensity of the external radiation field and the inverse of the
characteristic cloud column density.  Grains reach terminal velocity
on a timescale of $10^2$~yr, i.e. much shorter than the cloud
lifetime.  Thus any density enhancement can be expected to grow
linearly and the total dust enhancement can be expected to scale with
the total external radiation experienced by the cloud core over its
lifetime.  \citet{Seo2011} numerically simulated this effect,
including gas dynamics as well as coupling between the gas and dust,
and found that the dust concentration is enhanced by about an order of
magnitude in a narrow, inward-propagating shell, and depleted exterior
to that shell.  While the mean dust-to-gas ratio of a cloud is not
changed by migration, truncation of the cloud by photoevaporation or
internal collapse (to the exclusion of outer regions) would produce a
metal-enhanced object.  Both \citet{Whitworth2002} and \citet{Seo2011}
point out the relevance of this process to the metallicities of stars
and the formation of their planets.

A third scenario for dust-gas segregation involves accreting
protostars and their disks, or concomitant Bondi-Hoyle accretion from
the molecular cloud \citep{Throop2008}.  Photoevaporation of gas, but
not dust, from a disk produces an enhancement in the dust-to-gas ratio
which is inherited by the star as disk accretion continues.
Photoevaporation of disks driven by X-rays from the central star has
been proposed to explain the final, rapid stage of circumstellar disk
clearing \citep{Owen2012}.  Observations of blue-shifted lines of Ne
II \citep{Pascucci2009} and O I \citep{Hartigan1995} indicate that
heavy volatile elements as well as H and He are lost in these winds.
The temperature of the X-ray-heated ``surface'' of the disk from which
the winds flow is heated to a few thousand K \citep{Owen2012}, but
gravitational settling (and the formation of planetesimals) keeps dust
near the cooler mid-plane.  Disk masses are typically 1\% of the
central star, with a large scatter \citep{Andrews2013}; assuming the
convection zone of a young solar-mass star contains 0.1\msun,
accretion of an entire disk which has been severaly gas-depleted would
increase relative elemental abundances by $<0.04$ dex, or $<0.01$ dex
if the convection zone contains 0.4\msun.  This may fall short of 
explaining some of the observations.

Two important clues to the mechanism(s) of the observed variation are
the spatial scale on which the variation occurs, and the $T_C$ of the
``elbow'' in the relative abundance variation below which abundances
do not vary in any systematic manner.  Surveys of nearby, solar-age
field stars offer little information on the spatial scale of the
segregation as the stars are far from their birthplace.
\citet{Onehag2014} found no significant difference between M67 cluster
stars and the Sun, and proposed that the Sun also formed in a dense
cluster analogous to M67.  They suggested that the common pattern of
relative abundances was established by the removal of dust by
radiation drift.  But there are several problems with this
explanation.  Star formation does not take place in HII regions but in
surrounding neutral gas, and, as they point out, the expansion of an
HII region should outpace dust drift.  This explanation also requires
most stars to form after an earlier generation of massive stars that
move the dust.  It also requires that the vast majority of nearby
solar-type stars have {\it not} formed in dense stellar clusters,
something not supported by cluster statistics \citep{Williams2007}.
Finally, abundance differences between stars in physical pairs
requires gas-dust segregation on scales smaller than the cloud core.

The other clue is the value $T_C$ below which relative abundance
variation disappears, as this suggests the temperature and hence
location where the dust and gas segregation occurs.  If this ``knee''
in the abundance variation vs. $T_C$ is 1000~K, as many data sets
suggest \citep{Melendez2009}, this would seem to rule out a molecular
cloud setting with $T \sim 20-100$~K where even some C and O condense
as organics and ices.  All of these observations point to gas-dust
segregation during the formation or subsequent evolution of an
accretion disk as a plausible explanation of the abundance trends.  

\section{Discussion}
\label{sec.discussion}

Neither observations of M dwarfs nor models of Galactic chemical
evolution support the premise that there are a significant number of
stellar systems with primordial C/O $\sim 1$ in the solar
neighborhood, i.e. within the samples of Doppler radial velocity
surveys.  Moreover, if the process of planet formation is universal,
then studies of primitive meteorites show that the chemical
composition of small, rocky planets is controlled largely by the
composition and thermal processing of pre-existing dust from the
parent molecular cloud, rather than condensation from gas with the
stellar composition, and that this dust is likely to have a C/O
reflecting that of the bulk ISM.  Finally, small differences in the
relative patterns between the Sun and solar ``twins'' can best be
explained by dust-gas segregation at stellar scales and temperatures
of up to 1000~K, i.e. in accreting protostars, rather than planet
formation per se.

When comparing the Sun to the solar neighborhood it is usually assumed
that they share a common history.  However, it is also possible that
the two are not related.  \citet{Nieva2012} estimated ``cosmic''
standard abundances (rather the present-day abundances in the solar
neighborhood) using bright, slowly-rotating, early-type B stars.  Due
to their short main sequence lives, B stars contain abundance
information that is both contemparaneous and local, and their purely
radiative atmospheres are comparatively simple to model, although UV
photoionization rates must be correctly modeled \citep{Lyubimkov2013}.
\citet{Nieva2012} report metallicities that are very close to solar,
depsite 4.5~Gyr of intervening GCE, and a mean C/O of 0.37, lower than
the solar value of 0.55.  They explain both discrepancies by appealing
to outward migration of the Sun in the galactic disk.  These findings
warrant further investigation.

Depletion patterns in the ISM suggest that the C/O of dust largely
reflects the ISM; O is incorporated more rapidly in grains than C in
molecular clouds but it is depleted from grains more quickly in the
intercloud medium, and these two effects approximately balance.
Variation in the efficiency of grain erosion by SN shocks and UV
radiation could produce modest variations in the C/O, however.  While
the phases of O in interstellar grains are clearly water ice and
silicates, the phases of C are more controversial.  Volatility
patterns suggest C is {\it not} in refractory phase(s) like graphite
and carbides but instead more volatile organic molecules, in
particular aromatic hydrocarbons
\citep{Jones2009,Jones2013,Chiar2013}.  Aliphatic hydrocarbons have
been suggested as the source of the ubiquitious ISM absorbtion feature
at 3.4$\mu$m but the absence of polarization in this line is difficult
to reconcile with a scenario of condensation of carbonaceous mantles
onto silicate cores \citep{Li2014}.  Surface reactions could explain
why growth rate of C in grains is very slow with respect to other
elements.

While these considerations plus a bulk ISM C/O$<1$ are sufficient to
explain the oxidized nature of dust in the Solar System, it remains to
be explained why primitive meteorites in the Solar System are very
depleted in both C and O, especially C.  Based on the meteorite
abundances and associated uncertainties compiled by
\citet{Lodders2003} and solar abundances and uncertainties from
\citet{Caffau2011}, the depletion has a stochiometric ratio of $0.93
\pm 0.34$, i.e. consistent with unity and depletion by the formation
and removal of CO.  In this scenario the removal of C or O is not
controlled by volatility, otherwise either C or O would be far more
depleted than the other, depending on the effective $T_C$.  This can
occur if the reaction occurs in situ, i.e. at temperatures lower than
the $T_C$s of the C and O phases ($\sim 100$K) where both elements are
retained in grains, and/or where vapor is not lost by turbulent
mixing, i.e. in ``dead zone'' where turbulence is suppressed.

Primitive meteorites, considered analogs to the now-lost building
blocks of the Solar System, also include the highly-reduced enstatite
chondrites.  Enstatite chondrites contain carbide minerals but
virtually no water and have been explained by equilibration of solids
with a gas having C/O = 0.83 \citep{Grossman2008}.  Interestingly,
removal of an amount of C and O equal to that incorporated in CI
chondrites produces a residual gas with a C/O of 0.93.
\citet{Hutson2000} have proposed that removal of refractories from a
gas of solar composition and equilibration of solids with the
remaining gas could produce the precursors for enstatite chondrites.
If enstatite chondrites formed from recycling of the precursors of
terrestrial planets in an oxygen-poor gas it could explain why the
enstatite chondrites and Earth lie along the same mass-dependent
fractionation line in a three oxygen-isotope plot.  This equilibration
would necessarily have occurred at a comparatively low dust-to-gas
ratio and lower temperatures where moderately volatile elements such
as Fe would not be lost.

Al-Mg isotopes of CAIs in unequilibrated enstatite chondrites indicate
that these refractory inclusions formed in the same region as CAIs in
other chondrites, but were subsequently exposed to reducing conditions
\citep{Guan2000}.  Reducing conditions could have been established
sequentially or in a different part of the protoplanetary disk.  The
formation times of enstatite chondrites is an area of active research.
Dating of sulfides in unequilibrated E chondrites using the Fe-Ni and
Mn-Cr short-lived radionuclide chronometers gives ages of 12-13 Myr
after CAI formation \citep{Wadha1997}.  Re-analysis of the Mn-Cr
isotope data for a single sulfide in the MacAlpine Hills 88136 EL3
chondrite \citep{Guan2007,Telus2012} plus an initial
$^{53}$Mn/$^{55}$Mn of $5.1 \times 10^{-6}$ \citep{Yin2007} gives an
age of 10~Myr.

In summary, I propose the following scenario for the chemistry of rocky
planets:

\begin{itemize}
\item Planet formation proceeds from a two-component ISM (gas and
  dust) which are never fully equilibraxted during the formation process.
\item The composition of the precursor material of Solar System
  material is interstellar dust with refractory abundances and a C/O
  ratio approximately equal to the bulk ISM; this was set by grain
  growth in molecular clouds and destruction between clouds.
\item Planetesimals, as represented by carbonaceous (CI) chondrites, formed
  under the oxidizing, high dust-to-gas ratio conditions
  established by settling of interstellar grains to the disk mid-plane and
  subsequent depletion of C and O by formation and removal of CO.
\item Enstatite chondrites represent reduced material that
  equilibrated with gas after removal of solids in approximately
  CI proportions, possibly during a second generation of planetesimal
  formation.
\item The chemical composition of rocky exoplanets could be set by
  mixing of oxidized and reduced generations of planetesimals.  The
  mixing ratio could be determined by the effiency with which the
  first, oxidizing generation incorporated disk solids, as well as the
  dynamics of accretion in the disk.
\end{itemize}

Giant planets, unlike the small, rocky planets considered in this
work, accrete massive gas envelopes that would include volatile
species such as CO, and the considerations described above do not
preclude the possibility of C-rich atmospheres in giant planets.
Detections of such objects have been claimed, but are controversial
\citep[e.g.][]{Madhusudhan2011,Crossfield2012,Swain2013,Line2014,Stevenson2014,Hansen2014}.
If the core-first model of giant planet formation is correct, removal
of O as silicates and sequestration into a core increases the C/O
ratio of the gas that is subsequently captured into the planet's
envelope.  Beyond the ice-line, condensation of water and removal of
more O would drive the C/O of the gas even closer to unity
\citep{Oberg2011}.  Dissociation of CO in the disk and removal of the
O as water ice would further enhance this ratio, and indeed the
enrichment of planetary water in $^{17}$O and $^{18}$O relative to the
Sun is thought to be a signature of this process \citep{Clayton2002}.
These processes may facilitate the formation of reduced, enstatite
chondrite-like planetesimals as discussed above.

\acknowledgments

The author thanks Derek Homeier and France Allard for generating the
C/O=1 cases with the PHOENIX atmosphere model, Gary Huss for
discussions which stimulated this work, and Larry Nittler for
comments.  This research was supported by NASA grants NNX10AQ36G and
NNX11AC33G to EG.


\appendix
\section{Galactic Chemical Evolution Model}
\label{sec.gcemodel} 

I computed changes in the abundance of five isotopes in the vicinity
of the solar galactocetnric radius: \ctwo, \cthree, \osix, \oseven,
and \oeight.  The model accounts for the production of these isotopes
and their release into the ISM in SN explosions, winds from massive
stars, and AGB winds, incorporation of isotopes into long-lived
low-mass stars, and dilution of the ISM by the infall of metal-poor
gas.  The ISM is described by a two-box model with a lower-density,
warmer intercloud medium (ICM) which spawns molecular clouds, and a
giant molecular cloud (GMC) component that can form stars.  The mean
lifetime of molecular clouds is $\tau_{\rm GMC}$ and the residence
time of gas in the ICM before condensing into a cloud is $\tau_{\rm
  ICM}$.  The equations of motion for the mass surface density $m$ of
the two components are:
\begin{equation}
\frac{dm_{\rm ICM}}{dt} = - \frac{m_{\rm ICM}}{\tau_{\rm ICM}} + \frac{m_{\rm GMC}}{\tau_{\rm GMC}}  +  S_{\rm ICM} + S_{\rm AGB} + F,
\end{equation}
\begin{equation}
\frac{dm_{\rm GMC}}{dt} = \frac{m_{\rm ICM}}{\tau_{\rm ICM}} - \frac{m_{\rm GMC}}{\tau_{\rm GMC}} + S_{\rm GMC} - R\mathcal{M} ,
\end{equation}
where $F$ is the infall rate, $R$ is the rate of formation rate of
stars with $M > 1$\msun{}, $\mathcal{M}$ is the ratio of the total
stellar mass to the mass in stars $>1$\msun, $S_{\rm ICM}$ is the flux
of SN ejecta into the ICM from progenitors that explode in time $T >
\tau_{\rm GMC}$, $S_{\rm GMC}$ is the flux of SN ejecta into the
parent molecular cloud from progenitors that contribute to GMCs in
time $T < \tau_{\rm GMC}$, and $S_{\rm AGB}$ is the wind from AGB
stars, all contributing to the ICM.  The fluxes from AGB winds and SN
ejecta are given by:
\begin{equation}
\label{eqn.production}
S = \int_{M_1}^{M_2} \frac{R(T(M_*))}{\langle M \rangle} E(M_*) f(M_*) dM_*,
\end{equation}
where $\langle M \rangle$ is the mean mass of stars with $M_* >
1$\msun, $E$ is the ejected mass in winds and/or explosions, $f$ is
the fractional number of stars per unit mass in the IMF, and $M_1$ and
$M_2$ are the minimum and maximum masses contributing mass to the
three different cases, and $P$ is the mass in the wind or ejecta from
a progenitor of mass $M_*$.  The equations for the mass surface
densities of stars ($m_*$) and stellar remnants ($m_r$) are
\begin{equation}
\frac{dm_*}{dt} = R \mathcal{M} - \int_{M_1}^{M_2}R(T(M_*))f(M_*)dM_*,
\end{equation}
and 
\begin{equation}
\frac{dm_r}{dt} = \int_{M_1}^{M_2} \frac{R(T(M_*))}{\langle M \rangle} \left[M_*-P(M_*)\right] f(M_*) dM_*,
\end{equation}
where the limits of integation are over any stars that are moving off
the main sequence.  The star formation rate is related to the total
mass surface density of gas using a Schmidt-Kennicut law:
\begin{equation}
R = R_0 \left(\frac{m_{\rm GMC} + m_{\rm ICM}}{m_0}\right)^{\beta}
\end{equation}

The equations governing the the mass fraction of the $i$th isotope is:
\begin{equation}
m_{\rm ICM}\frac{dX^i_{\rm ICM}}{dt} = \frac{m_{\rm GMC}X_{GMC}^i}{\tau_{\rm GMC}} - \frac{m_{\rm ICM}X_{\rm ICM}^i}{\tau_{\rm ICM}} + S^i_{\rm ICM} + S^i_{\rm AGB} + F X^i_{0},
\end{equation}
\begin{equation}
m_{\rm GMC}\frac{dX^i_{\rm GMC}}{dt} = -\frac{m_{\rm GMC}X_{\rm GMC}^i}{\tau_{\rm GMC}} + \frac{m_{\rm ICM}X_{\rm ICM}^i}{\tau_{\rm ICM}} + S^i_{\rm GMC} - R \mathcal{M}X^i_{\rm GMC},
\end{equation}
$X^i_0$ are the isotopic abundances of the infalling gas, and 
\begin{equation}
\label{eqn.iso_production}
S_i = \int_{M_1}^{M_2} \frac{R(T(M_*))}{\langle M \rangle} P^i(M_*) f(M_*) dM_*,
\end{equation}
where $P^i$ is the production (equal to the nucleosynthetic yield plus
the original mass of the isotope).  The C/O ratio is calculated by
summing the appropriate mass fractions divided by the atomic weights.
An approximate metallicity is also calcated from the sum of all the
isotopes relative to the abundances in the Sun.

I considered nucleosynthesis in the progenitors of AGB stars
(1-6.5\msun), ``super-AGB'' stars (6.5-11\msun), and SN (11-120\msun).
I used the AGB yields from \citet{Karakas2010} for AGB progenitors up
to 6.5\msun.  For super-AGB progenitors with masses between 7 and
11\msun{} I used the productions from \citet{Doherty2014}, or
\citet{Siess2010} otherwise.  For SN yields I adopted the values in
\citet{Kobayashi2006} and \citet{Kobayashi2011} for regular SN with
energies of $10^{51}$ erg and their re-run calcalations for the cases
of $M_* = 18$\msun{} and $Z = 0.004$, and $M_* = 25$ and $Z = 0.02$.
These were supplemented with values from \citet{Portinari1998} for
progenitor masses of 12, 60, 100, and 120\msun.  In order to arrive at
a solar C/O with reasonable choices of parameters I find it necessary
to include yields from the {\it winds} of massive stars with $M_* >
40$\msun{} estimated by \citet{Portinari1998} (Table 3 in that work).

\begin{figure}
\includegraphics[width=84mm]{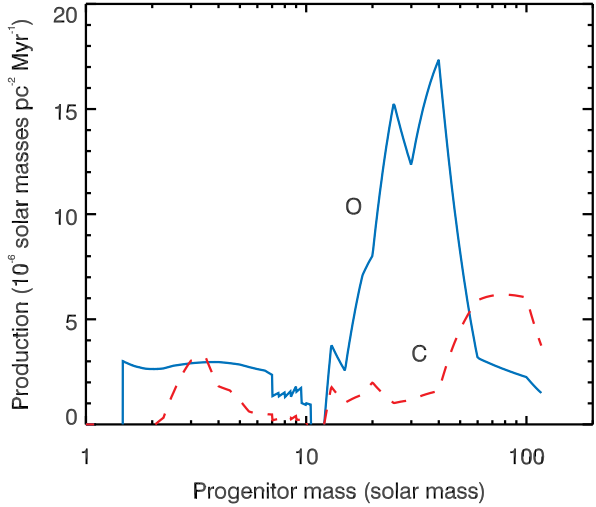}
\caption{Predicted mass of oxygen (blue, solid) and carbon (red,
  dashed) from stars vs. progenitor mass for a solar-metallicity
  population weighted by an IMF with power-law index $\alpha = 2.6$}
 \label{fig.yield}
\end{figure}

The literature for SN and AGB nucleosynthesis is very heterogeneous,
with calculations performed for different ranges/values of progenitor
masses and metallicities, and it was necessary to interpolate or
extrapolate for some values.  Specifically, to estimate the production
of 6.5\msun{} AGB stars with sub-solar progenitor metallicities I
scaled the \citet{Karakas2010} values for sub-solar metallicity
6\msun{} progenitors by the ratio of the 6.5\msun{} to 6\msun{}
production for solar-metallicity.  For productions from super-AGB
stars with progenitor masses outside the caclulated range, I either
used the production from the most massive progenitor with the same
metallicity, or linearly interpolated between the values for the most
massive AGB and least massive super-AGB progenitors with the same
metallicity.  For a review of the many parameters and uncertainties
that enter these calculations see \cite{Karakas2014}.  I linearly
interpolated the production onto a grid of 1000 masses over
1-120\msun{} with intervals chosen such that the IMF has equal total
mass in each bin.  Figure \ref{fig.yield} shows the calculated yields
(production - initial incorporation) for solar metallicity.  These
values were then used to evaluate Eqns. \ref{eqn.production} and
\ref{eqn.iso_production}.

I searched for the parameter values that best reproduce the
observational constraints using Monte-Carlo Markov Chain analysis.
The constraints are the age of the Sun, the present mass surface
densities of stars, stellar remnants, and total gas at the solar
galactocentric radius, solar C/O, the current metallicity of the ISM
(taken to be +0.14 based on the metallicities of M44 and the Hyades),
and the mean (-0.05) and intrinsic standard-deviation (0.18 dex) of
the metallicity distribution in the solar neighborhood
\citep{Gaidos2014}.  I adopted standard errors of 0.12 in C/O
\citep{Caffau2011}, 1.5, 1, and 1 \msun~pc$^{-2}$ for stars, remnants,
and gas, 50~Myr for the absolute age of the Sun, and 0.01 and 0.02 dex
in the mean and standard deviation of the present metallicity.  I
varied the exponential infall timescale, the metallicity of the
infalling material, the age of the Galactic disk in the solar annulus,
the high-mass IMF index $\alpha$, and the Schmidt-Kennicut index
$\beta$.

\begin{figure}
\includegraphics[width=84mm]{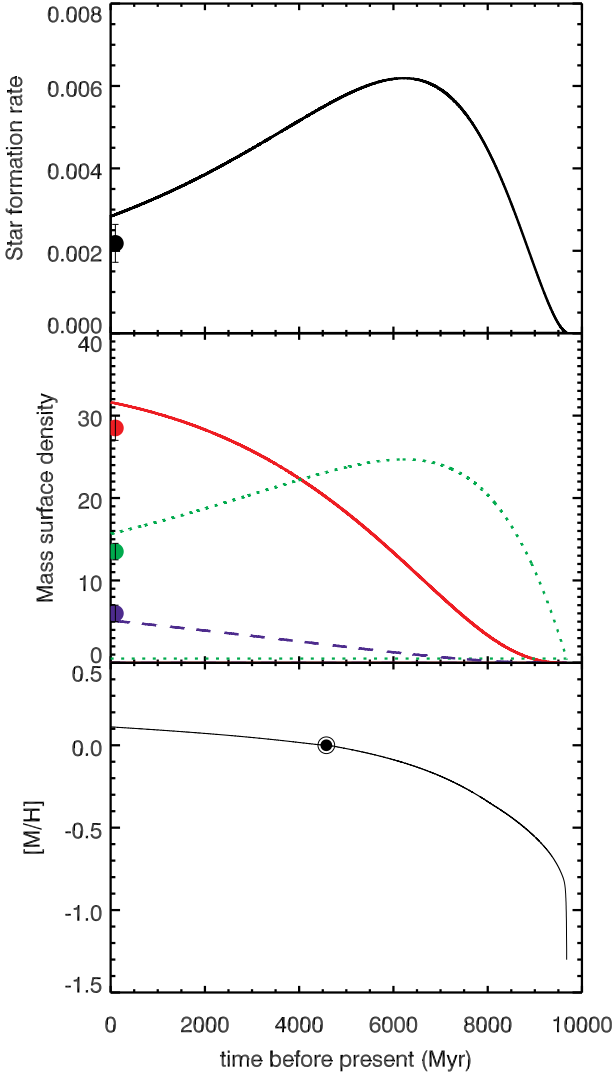}
\caption{Predictions of the GCE model compared to the Sun and values
  at present.  Top: star formation rate per area compared to the
  present estimated value.  Middle: mass surface densities of stars
  (solid, red), gas (green, dotted), and stellar remnants (blue,
  dashed) compared to estimates of present values.
  \citep[points,][]{Naab2006}.  Bottom: metallicity, here taken to be
        [C+O/H].}
 \label{fig.gce}
\end{figure}

The simulation is able to adequately reproduce the observed properties
of the Galactic disk in the solar neighborhood ($\chi^2 = 35$ with
$\nu = 3$).  The most significant deviation is the over-prediction of
the present ISM gas mass and underprediction of the current
metallicity.  The adopted or fit values for the simulation are given
in Table \ref{tab.params}.  The predicted evolution of the SFR, mass
surface densities, and the metallicity of the ISM are plotted in
Fig. \ref{fig.gce}.  Estimates of the current total star formation rate
of the Milky Way cluster around $1.9 \pm 0.4$~\msun~yr$^{-1}$
\citep{Chomiuk2011}.  Presuming that star formation has an exponential
radial distribution with scale length of 3.5~kpc \citep{Wolfire2003}
then the rate at the solar radius will be about $2.7 \pm 0.6 \times
10^{-3}$~\msun~Myr$^{-1}$~pc$^{-2}$.  This compares favorably with the
predicted value at the present (Fig. \ref{fig.gce}).

\begin{table*}[h]
\centering
\begin{minipage}{110mm}
\caption{GCE Simulation Parameters}
\label{tab.params}
\begin{tabular}{@{}lllll@{}}
\multicolumn{1}{c}{Symbol} & \multicolumn{1}{c}{Parameter} & \multicolumn{1}{c}{Value} & \multicolumn{1}{c}{Units} & \multicolumn{1}{c}{Reference} \\
\hline
\multicolumn{5}{c}{Solar Parameters} \\
\hline
$X^{\rm O}$ & Solar oxygen & $6.73 \times 10^{-3}$ & --- & \citet{Caffau2011} \\
$X^{\rm C}$ & Solar carbon & $2.73 \times 10^{-3}$ & --- & \citet{Caffau2011} \\
\hline
\multicolumn{5}{c}{Stellar Parameters} \\
$\alpha$ & IMF index & 2.61 & --- & Fit\\
$\mathcal{M}$ & Low mass/high-mass ratio & 2.00 & --- & Fit\\
\hline
\multicolumn{5}{c}{Galactic Parameters} \\
$m_*(T)$ & Present  stellar mass density & $28.5 \pm 1.5$ & \msun~pc$^{-2}$ & \citet{Naab2006}\\
$m_g(T)$ & Present gas mass density & $13.5 \pm 1$ & \msun~pc$^{-2}$ & \citet{Naab2006}\\
$m_r(T)$ & Present remnant mass density & $6 \pm 1$ & \msun~pc$^{-2}$ & \citet{Naab2006}\\
$\tau_{\rm GMC}$ & GMC gas residence time & 30 & Myr & \\
$\tau_{\rm IMC}$ & IMC gas residence time & 150 & Myr & \\
$T_{\rm disk}$ & Age of disk & 9554 & Myr & Fit\\
$\tau_{\rm infall} $ & Infall e-folding time & 2630 & Myr & Fit \\
$[M/H]_0$ & Metallicity of infalling gas & -1.3 & --- & Fit \\
$\beta$ & SFR rate index & 1.71 & --- & Fit\\
\hline
\end{tabular}
\medskip
\end{minipage}
\end{table*}
\end{document}